\documentclass[%
 aip,
rsi,%
 amsmath,amssymb,
 reprint,%
]{revtex4-1}

\usepackage{color}
\usepackage{graphicx}
\usepackage{dcolumn}
\usepackage{bm}
  \usepackage{algorithm} 
  \usepackage{algpseudocode}  
\usepackage{tikz}
\usetikzlibrary{shapes.geometric, arrows}

\begin{document}

\preprint{AIP/123-QED}

\title{
%
Speeding-up {\it Ab Initio} Molecular Dynamics with Hybrid Functionals using Adaptively Compressed Exchange Operator based Multiple Timestepping 
%
%
}

\author{Sagarmoy Mandal}
\affiliation{ 
Department of Chemistry, Indian Institute of Technology Kanpur, Kanpur - 208016, India
}%

\author{Nisanth N. Nair}
\email{nnair@iitk.ac.in}

\affiliation{ 
Department of Chemistry, Indian Institute of Technology Kanpur, Kanpur - 208016, India
}%

\date{\today}

\begin{abstract}
{\it Ab initio} molecular dynamics (AIMD) simulations using hybrid density functionals and plane waves
are of great interest owing to the accuracy of this approach in treating condensed matter systems.
On the other hand, such AIMD calculations are not routinely carried out 
since the computational
cost involved in applying the Hartree Fock exchange operator is very high.
%
%
%
In this work, we make use of a strategy that combines adaptively compressed exchange operator formulation and multiple time step integration  to significantly reduce the computational cost of these simulations. 
%
%
We demonstrate the efficiency of this approach for a realistic condensed matter system.
%
%
%
%
\end{abstract}

\maketitle

%
{\it Ab initio} molecular dynamics (AIMD) simulations with density functional theory (DFT) and plane wave (PW) basis set are the methods of choice in studying structural 
and dynamic properties of condensed matter systems.\cite{marx-hutter-book}
Usage of density functionals at the level of Generalized Gradient Approximation (GGA) is commonplace for these simulations because more than 
a million energy and force evaluations are computationally achievable
by taking advantage of parallel programs and parallel computing platforms.
%
Contrarily, hybrid density functionals are preferred over GGA functionals for improved accuracy in AIMD simulations.\cite{JPCB_AIMD_HFX,JCTC_AIMD_HFX,JCP_AIMD_HFX,JPCB_water_hfx}
Computations of energy and gradients at the hybrid functional level using PW basis set have prohibitively high computational cost resulting from the 
application of the exact exchange operator on each of the occupied orbitals.
One of the ways to increase the efficiency of such AIMD simulations is by making use of multiple time step (MTS) algorithms\cite{MTS_1,r-RESPA} among others.\cite{PRB_Car_Wannier,JCP_AIMD_HFX,JCTC_RSB,JCTC_RSB_1,HFX_Goedecker,JCP_sagar} 
%
%
%
%
In this respect, the reversible reference system propagator algorithm (r-RESPA)\cite{r-RESPA} has been used by several authors.\cite{HFX_Hutter_JCP,MTS_AIMD_Ursula,MTS_AIMD_Steele_3}
In the r-RESPA MTS approach, artificial time scale separation in the ionic force components due to computationally intensive Hartree Fock exchange (HFX) contribution and the computationally cheaper rest of the terms is made.\cite{HFX_Hutter_JCP,MTS_AIMD_Ursula} 
%
In this manner, MTS scheme allows us to compute HFX contributions less frequently 
compared to the rest of the contributions to the force, thereby reducing the 
overall computational cost in performing AIMD simulations.
%

Here we propose a new way to take advantage of the r-RESPA scheme for
performing AIMD using hybrid functionals and PWs.
This scheme is based on the recently developed adaptively compressed exchange (ACE) operator approach.\cite{ACE_Lin,ACE_Lin_1}
%
We exploited some property of the ACE operator to artificially split the ionic forces into fast and slow.
%
%

%
The self consistent field (SCF) solution of hybrid functional based Kohn-Sham (KS) DFT equations requires application of the exchange operator ${\mathbf V}_{\rm X}=-\sum_{j}^{N_{\rm orb}} \frac{ | \psi_{j} \rangle  \langle \psi_{j} |}{r_{12}}$ on each of the KS orbitals $|\psi _{i} \rangle$:
\begin{equation}
{\mathbf V}_{\rm X}|\psi _{i}\rangle=- \sum_{j}^{N_{\rm orb}} |\psi _{j} \rangle \left \langle\psi _{j} \left | \left ( r_{12}\right )^{-1} \right | \psi _{i}\right \rangle , \enspace ~~i=1,....,N_{\rm orb} \enspace .
\end{equation}
Here, $N_{\rm orb}$ is the total number of occupied orbitals.
The evaluation of $\left \langle\psi _{j} \left | \left ( r_{12}\right )^{-1} \right | \psi _{i}\right \rangle$ is usually done in reciprocal space\cite{JCP_HFX_Voth,PRB_Car_Wannier} using Fourier transform (FT).
%
If $N_{\rm G}$ is the total number of PWs, the computational cost for doing FT scales as $N_{\rm G}\log  N_{\rm G}$ on using fast Fourier transform (FFT) algorithm.
The total computational cost  scales as $N_{\rm orb}^2 N_{\rm G}\log  N_{\rm G}$,\cite{JCP_HFX_Voth} as operation of ${\mathbf V}_{\rm X}$ on all the KS orbitals requires $N_{\rm orb}^{2}$ times evaluation of $\left \langle\psi _{j} \left | \left ( r_{12}\right )^{-1} \right | \psi _{i}\right \rangle$. 
%

In the recently developed ACE operator formulation,\cite{ACE_Lin} the full rank ${\mathbf V}_{\rm X}$ operator is approximated by the ACE operator ${\mathbf V}_{\rm X}^{\rm ACE}= -\sum_{k}^{N_{\rm orb}}  | P_{k} \rangle  \langle P_{k} |$ using a low rank decomposition.
Here, $\{|P_{k} \rangle\}$ is the set of ACE projection vectors which can be computed through a series of simpler linear algebra operations.
%
Now, the evaluation of the action of ${\mathbf V}_{\rm X}^{\rm ACE}$ operator on KS orbitals can be done with $N_{\rm orb}^{2}$ number of simpler inner products as
\begin{equation}
{\mathbf V}_{\rm X}^{\rm ACE}|\psi _{i}\rangle=- \sum_{k}^{N_{\rm orb}} |P_{k} \rangle  \left \langle P_k | \psi_{i} \right \rangle , \enspace ~~i=1,....,N_{\rm orb}  \enspace .
\end{equation}
The advantage of the ACE approach is that the cost of applying the ${\mathbf V}_{\rm X}^{\rm ACE}$ operator on each KS orbitals is much less as compared to ${\mathbf V}_{\rm X}$ operator.
At the first SCF step, ${\mathbf V}_{\rm X}^{\rm ACE}$ operator can be constructed through the computation of $\{{\mathbf V}_{\rm X}|\psi _{i}\rangle \}$, which is the costliest step ($N_{\rm orb}^{2}$ times evaluation of $\left \langle\psi _{j} \left | \left ( r_{12}\right )^{-1} \right | \psi _{i}\right \rangle$).
As HFX has only a minor contribution to the total energy, an approximate energy computation
is possible by using the previously constructed ${\mathbf V}_{\rm X}^{\rm ACE}$ operator without updating it for the rest of the SCF iterations.
It is again stressed that, once the ${\mathbf V}_{\rm X}^{\rm ACE}$ operator is constructed, its low rank structure allows the easy computation of $\{{\mathbf V}_{\rm X}^{\rm ACE}|\psi _{i}\rangle \}$ in the subsequent SCF iterations.
We exploit this property of the ACE operator to combine with the r-RESPA scheme. 
%

%
In the r-RESPA method,\cite{r-RESPA} symmetric Trotter factorization of the classical time evolution operator is carried out.
Let that ionic force can be decomposed into slow and fast components as $F_K= F_K^{\rm fast}+ F_K^{\rm slow}$, $K=1,\cdots,3N$, for a system containing $N$ atoms.
In this case, the Liouville operator $L$ 
can be written as
\begin{equation}
iL= iL^{\rm fast}_1 + iL^{\rm fast}_2 + iL^{\rm slow} \enspace,
\end{equation}
with
\begin{equation}
 iL^{\rm fast}_1=  \sum_{K=1}^{3N}\left [ \dot{X}_K \frac{\partial}{\partial {X}_K}  \right ]  ,~~ iL^{\rm fast}_2=  \sum_{K=1}^{3N}\left [  F_{K}^{\rm fast} \frac{\partial}{\partial {P}_K} \right ]
\end{equation}
and
\begin{equation}
iL^{\rm slow} = \sum_{K=1}^{3N}\left [ F_{K}^{\rm slow} \frac{\partial}{\partial {P}_K} \right ] \enspace.
\end{equation}
Here, $\{X_K\}$ and $\{P_K\}$ are the Cartesian coordinates and the conjugate momenta of the particles.
Using symmetric Trotter factorization, we arrive at 
%
\begin{equation}
\begin{split}
%
\exp (& iL \Delta t) \approx \exp \left (iL^{\rm slow} \frac{\Delta t}{2} \right) \\ &  \times \left [ \exp \left (iL^{\rm fast}_2 \frac{\delta t}{2} \right)   \exp \left(iL^{\rm fast}_1\delta t \right)  \exp \left(iL^{\rm fast}_2 \frac{\delta t}{2} \right)  \right ]^n \\ & \times  \exp \left (iL^{\rm slow} \frac{\Delta t}{2} \right) \enspace. 
%
\end{split}
\end{equation}
%
%
%
Here, the large time step $\Delta t$ is chosen according to the time scale of variation of slow forces ($\{F_K^{\rm slow}\}$)
and the smaller time step $\delta t=\Delta t/n$ is chosen according to the time scale of fast forces ($\{F_K^{\rm fast}\}$).
%
%
%

\begin{figure}
\includegraphics[scale=0.37]{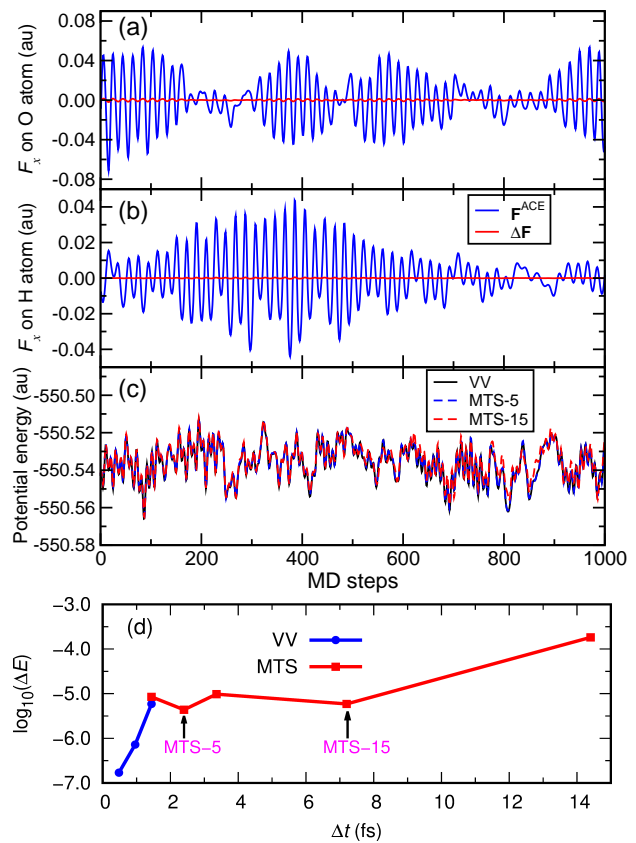}
\caption{\label{f_component}
Test results for 32-water system using PBE0 functional:
One of the components of $\mathbf F^{\rm ACE}$ and $\Delta \mathbf F$ on an arbitrarily chosen (a) oxygen and (b) hydrogen atoms;
(c)  Comparison of potential energy during {\bf VV}, {\bf MTS-5} and {\bf MTS-15} simulations in NVE ensemble; 
%
%
(d)  $\log_{10}(\Delta E)$ for different $\Delta t$ values in {\bf VV} and {\bf MTS} simulations calculated from 5~ps long trajectories. 
}
\end{figure}

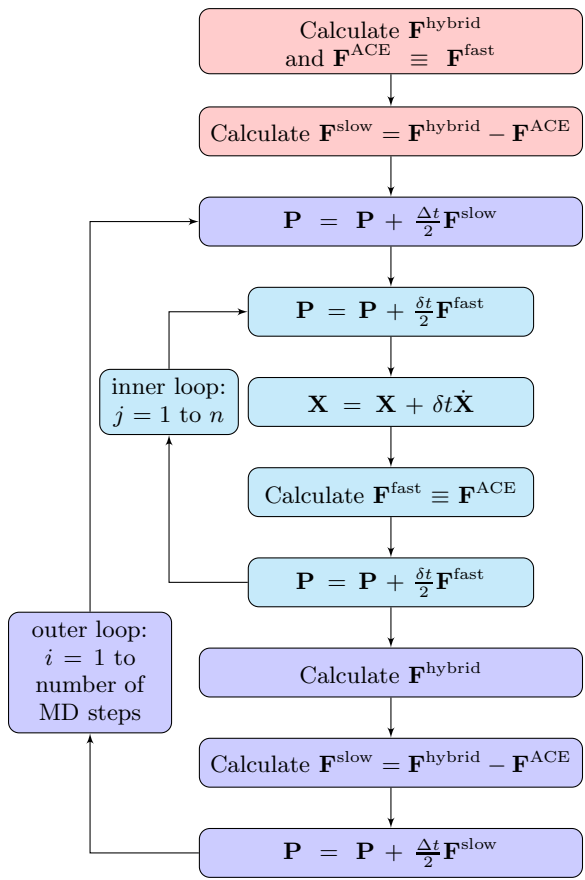
\begin{figure}
\centering
\tikzstyle{decision} = [diamond, draw, fill=blue!20, text width=7em, text badly centered, node distance=2.5cm, inner sep=0pt]
\tikzstyle{block} = [rectangle, draw, fill=blue!20, text width=15em, text centered, rounded corners, minimum height=2em]
\tikzstyle{block1} = [rectangle, draw, fill=cyan!20, text width=11em, text centered, rounded corners, minimum height=2em]
\tikzstyle{line} = [draw, -latex']
\tikzstyle{cloud} = [draw, ellipse,fill=red!20, node distance=2.5cm, minimum height=1em]
     
\begin{tikzpicture}[node distance = 1.5cm, auto]
\node[block, fill=red!20, node distance=1.2cm](for0){Calculate $\mathbf F^{\rm hybrid}$ and $\mathbf F^{\rm ACE} \equiv \mathbf F^{\rm fast}$};
\node[block, fill=red!20, below of=for0, node distance=1.2cm](for01){Calculate $\mathbf F^{\rm slow}=\mathbf F^{\rm hybrid}-\mathbf F^{\rm ACE}$};
\node[block, below of=for01, node distance=1.2cm](mom1){$\mathbf P=\mathbf P+\frac{\Delta t}{2}\mathbf F^{\rm slow}$};
\node[block1, below of=mom1, node distance=1.2cm](mom2){$\mathbf P=\mathbf P+\frac{\delta t}{2}\mathbf F^{\rm fast}$};
\node[block1, below of=mom2, node distance=1.2cm](pos){$\mathbf X=\mathbf X+\delta t \dot{\mathbf X}$};
\node[block1, below of=pos, node distance=1.2cm](for){Calculate $\mathbf F^{\rm fast}\equiv \mathbf F^{\rm ACE}$};
\node[block1, left of=pos, node distance=2.95cm, text width=5em](i_loop){inner loop:  $j=1$ to $n$};
\node[block1, below of=for, node distance=1.2cm](mom3){$\mathbf P=\mathbf P+\frac{\delta t}{2}\mathbf F^{\rm fast}$};
\node[block, below of=mom3, node distance=1.2cm](for13){Calculate $\mathbf F^{\rm hybrid}$};
\node[block, below of=for13, node distance=1.2cm](for1){Calculate $\mathbf F^{\rm slow}=\mathbf F^{\rm hybrid}-\mathbf F^{\rm ACE}$};
\node[block, below of=for1, node distance=1.2cm](mom4){$\mathbf P=\mathbf P+\frac{\Delta t}{2}\mathbf F^{\rm slow}$};
\node[block, left of=for13, node distance=4.0cm, text width=6em](o_loop){outer loop: $i=1$ to number of MD steps};

\path [line] (for0) -- (for01);
\path [line] (for01) -- (mom1);
\path [line] (mom1) -- (mom2);
\path [line] (mom2) -- (pos);
\path [line] (pos) -- (for);
\path [line] (for) -- (mom3);
\path [line] (mom3) -- (for13);
\path [line] (for13) -- (for1);
\path [line] (for1) -- (mom4);
\path [line] (mom4) -| (o_loop);
\path [line] (o_loop) |- (mom1);
\path [line] (mom3) -| (i_loop);
\path [line] (i_loop) |- (mom2);
 
\end{tikzpicture}
\caption{Flowchart of the MTS propagation scheme 
proposed in this work.} \label{mts}
\end{figure}

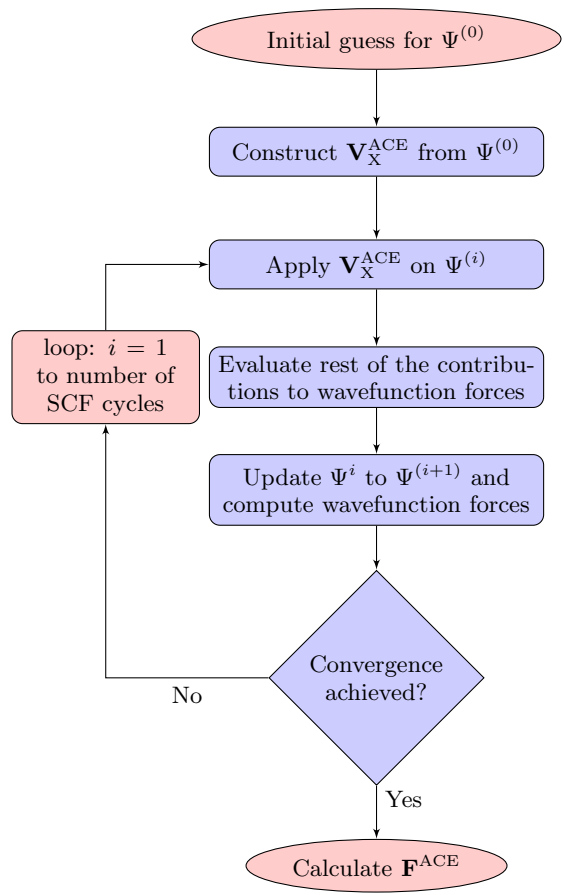
\begin{figure}
\centering
\tikzstyle{decision} = [diamond, draw, fill=blue!20, text width=7em, text badly centered, node distance=2.5cm, inner sep=0pt]
\tikzstyle{block} = [rectangle, draw, fill=blue!20, text width=13em, text centered, rounded corners, minimum height=2em]
\tikzstyle{line} = [draw, -latex']
\tikzstyle{cloud} = [draw, ellipse,fill=red!20, node distance=2.5cm, minimum height=1em]
     
\begin{tikzpicture}[node distance = 1.5cm, auto]
\node [cloud, text width=10em, text badly centered] (init) {Initial guess for $\Psi^{(0)}$};
\node [block, below of=init] (ace) {Construct ${\mathbf V}_{\rm X}^{\rm ACE}$ from $\Psi^{(0)}$};
\node [block, below of=ace] (apply) {Apply ${\mathbf V}_{\rm X}^{\rm ACE}$ on $\Psi^{(i)}$};
\node [block, below of=apply] (evaluate) {Evaluate rest of the contributions to wavefunction forces};
\node [block, left of=evaluate, node distance=3.6cm, text width=7em, fill=red!20] (loop) {loop: $i=1$ to number of SCF cycles};
\node [block, below of=evaluate] (update) {Update $\Psi^i$ to  $\Psi^{(i+1)}$ and compute wavefunction forces};
\node [decision, below of=update] (convergence) {Convergence achieved?};
\node [cloud, below of=convergence] (forces) {Calculate $\mathbf F^{\rm ACE}$};
 

\path [line] (init) -- (ace);
\path [line] (ace) -- (apply);
\path [line] (apply) -- (evaluate);
\path [line] (evaluate) -- (update);
\path [line] (update) -- (convergence);
\path [line] (convergence) -- node [near start] {Yes} (forces);
\path [line] (convergence) -| node [near start] {No} (loop);
\path [line] (loop) |- (apply);
 
\end{tikzpicture}
\caption{Flowchart showing the steps involved in computing $\mathbf F^{\rm ACE}$.} \label{fast_forces}
\end{figure}

%
%
%
%
Now, we split the contribution of ionic forces from the HFX part as
\begin{equation}
    F^{\rm hybrid}_K= F^{\rm ACE}_K+ \Delta F_K  \enspace , \enspace K=1,\cdots,3N \enspace 
\end{equation}
with $\Delta F_K = \left ( F^{\rm hybrid}_K -  F^{\rm ACE}_K \right )$.
Here, $\mathbf F^{\rm hybrid}$ is the ionic force computed with the full rank exchange operator ${\mathbf V}_{\rm X}$.
The term $\mathbf F^{\rm ACE}$ is the ionic force calculated using the low rank ${\mathbf V}_{\rm X}^{\rm ACE}$ operator.
%
%
%
In Figures~\ref{f_component}(a) and (b) we have shown the components of the $\mathbf F^{\rm ACE}$ and $\Delta \mathbf F$ for a realistic molecular system, where
$\mathbf V_{\rm X}^{\rm ACE}$ is calculated once at the beginning of a SCF while kept fixed during the remaining SCF cycles.
%
%
The clear difference in the time scale at which the two forces are varying
allowed us to use  $\mathbf F^{\rm ACE}$ as the fast ionic force and $\Delta \mathbf F$ as the slow ionic force, and combine it with the r-RESPA algorithm.
Here, the longer time step $\Delta t$ is chosen according to the time scale of variation of the computationally costly slow forces,
%
whereas the smaller time step $\delta t$ is taken as per the time scale of fast forces that are cheaper to compute.
%
In this way, we get the required speed-up using r-RESPA scheme to perform hybrid functional based AIMD simulations.
A flowchart of the method is given in Figure~\ref{mts} and~\ref{fast_forces}.
%

%
%

%
Benchmark calculations were carried out for a 32 water system where the molecules were taken in a supercell of dimensions 9.85~{\AA}$\times$9.85~{\AA}$\times$9.85~{\AA} with
 water density $\sim$1~g~cm$^{-3}$.
%
Calculations were carried out employing
the {\tt CPMD} program\cite{cpmd} where the proposed method has been implemented.
%
The PBE0\cite{JCP_PBE0_model} exchange correlation functional was employed together with the 
norm-conserving Troullier-Martin type pseudopotentials.\cite{PRB_TM}
A PW cutoff energy of 80~Ry was used.
Born-Oppenheimer molecular dynamics (BOMD) simulations were carried out to perform MD simulations at the microcanonical (NVE) and canonical (NVT) ensembles.
%
%
In order to perform canonical ensemble AIMD simulation, we employed Nos{\'e}--Hoover chain thermostats\cite{NHC} and the temperature of the system was set to 300~K.
Addition of thermostats also helps to eliminate any resonance effects originated with the use of large time step.\cite{mts_resonance}
%
%
At every MD steps, wavefunctions were converged till the magnitude of maximum wavefunction gradient reached below $1\times 10^{-6}$ au.
The initial guess for the wavefunctions at every MD step was obtained using Always Stable Predictor Corrector Extrapolation scheme\cite{JCC_ASPC} of order 5.

%

%
%
\begin{figure}
\includegraphics[scale=0.5]{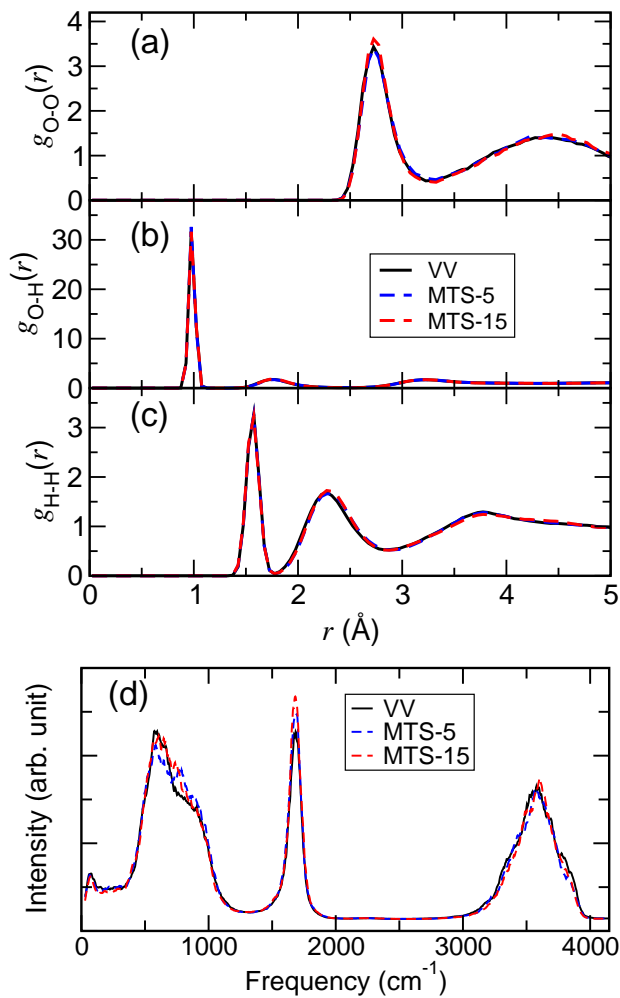}
\caption{\label{g_of_r} Radial distribution functions (RDFs) for bulk water simulation
from {\bf VV}, {\bf MTS-5} and {\bf MTS-15} trajectories at the level of PBE0: (a) O-O, (b) O-H,  and (c) H-H.
(d) Power spectrum of the same system computed from {\bf VV}, {\bf MTS-5} and {\bf MTS-15} trajectories.}
\end{figure}
%
%
\begin{table}
\caption{\label{table} Comparison of various quantities for {\bf VV}, {\bf MTS-5} and {\bf MTS-15} simulations in NVE ensemble.}
\begin{ruledtabular}
\begin{tabular}{lcccc}
Method & $\log_{\rm 10}(\Delta E)$\footnote{Calculated using Equation~\ref{energy_con} over  5~ps long trajectories.} & ${\Delta U}$/(au)\footnote{The average absolute deviation of potential energy in {\bf MTS-n} runs from the {\bf VV} run: $\Delta U=\left \langle \left | U^{\textbf {VV}} - U^{\textbf { MTS-n}} \right | \right \rangle$. Here, $U^{\textbf {VV/MTS-n}}$ is the potential energy at any time during {\bf VV}/{\bf MTS-n} run. This average is calculated over 1000 MD steps.} & $t_{\rm CPU}/(s)$\footnote{Average computational time per MD step (averaged over 500 MD steps) performed using identical 120 processors.} & speed-up\footnote{Speed-up is the ratio of $t_{\rm CPU}$ for {\bf VV} and {\bf MTS-n} runs.} \\
\hline
{\bf VV} & -6.8 & 0.0 & 258 & 1  \\
{\bf MTS-5} & -5.4 & 5.9$\times 10^{-4}$ & 64 & 4 \\
{\bf MTS-15} & -5.2 & 1.9$\times 10^{-3}$ & 38 & 7 \\
\end{tabular}
\end{ruledtabular}
\end{table}
\def\arraystretch{1.5}
\begin{table}
\caption{\label{table2} CPU time for various stages of the program.}
\begin{tabular}{|p{6cm}|c|}  \hline \hline 
CPU time per SCF using ${\mathbf V}_{\rm X}$ operator & 24~s \\ \hline
CPU time per SCF using ${\mathbf V}_{\rm X}^{\rm ACE}$ operator & 0.1~s  \\ \hline 
Average CPU~time for the construction of ${\mathbf V}_{\rm X}^{\rm ACE}$
at the beginning of every MD step & 24~s \\ \hline \hline  
\end{tabular}
\end{table}

To benchmark our implementation, we first compared the fluctuations in total energy 
using conventional velocity Verlet ({\bf VV}) integrator and MTS runs ({\bf MTS-n})
with $n=\Delta t/\delta t$, and $\delta t \approx 0.5$~fs for 32-water in a periodic box treated by PBE0 functional.
%
%
%
%
The magnitude of the total energy ($E$) fluctuations is measured by 
%
%
\begin{equation}
\label{energy_con}
\Delta E= \left < \left | \frac{E-\left \langle E \right \rangle}{\left \langle E \right \rangle} \right |  \right >   \enspace ,
\end{equation}
where $\left < \cdots \right >$ specifies time average.
%
%
%
In the case of {\bf VV} runs, 
$\log_{10}(\Delta E)$
increases with higher $\Delta t$ corresponding to the increase in total energy fluctuations as shown in Figure~\ref{f_component}(d).
We also observed that the use of a timestep greater than 1.4~fs 
in {\bf VV} runs leads to unstable trajectories with breaking of O-H covalent bonds.
In {\bf MTS-n} runs, we kept the inner timestep $\delta t$ fixed at 
$\sim$0.5~fs 
and varied outer timestep $\Delta t=n \, \delta t$.
%
The quality of the energy conservation in these runs depends on the value of $n$, which determines how large the outer timestep is compared to the inner timestep.
It is clear from Figure~\ref{f_component}(d) that
{\bf MTS-n} runs with $n$ up to 15 have total energy conservation comparable to 
 {\bf VV} run using a timestep $1.4$~fs.
%
Although the {\bf MTS-30} run (with 
$\Delta t=14.4$~fs) was showing higher total energy fluctuation, it was able 
to generate stable MD trajectories.
%
Notably, we observed good accuracy in {\bf MTS} runs with $n=15$ (i.e. {\bf MTS-15}); (see Table~\ref{table}).
%

%
In order to show the correctness of our proposed MTS scheme, we compared the fluctuation in potential energy for {\bf VV}, {\bf MTS-5} and {\bf MTS-15} runs for a short initial time period for the 32-water system (before the trajectories deviate due to growing numerical differences) in Figure~\ref{f_component}(c); see also Table~\ref{table}.
All these simulations were started with the same initial conditions.
%
We find that potential energy computed from the {\bf MTS-5} and {\bf MTS-15} trajectories 
are closely following the potential energy from the {\bf VV} run. 
%
%
%

%
As next, we carried out NVT simulations for the same system and computed static and dynamical properties of bulk water.
%
In particular, we calculated partial radial distribution functions (RDFs) and the power spectrum; see Figure~\ref{g_of_r}.
It is clear that the RDFs from the {\bf MTS} simulations are in excellent agreement with those from the {\bf VV} run (Figure~\ref{g_of_r} a,b and c).
Also, the power spectrum computed from these calculations are in excellent agreement (Figure~\ref{g_of_r} d).
Thus, we conclude that our {\bf MTS} scheme gives accurate description of the structural and dynamical properties.

We now compare the average computational time per MD step ($t_{\rm CPU}$) for {\bf MTS-n} and {\bf VV} runs; see Table~\ref{table} and~\ref{table2}.
We have achieved a speed-up of $\sim$4 fold for the 32-water system with {\bf MTS-5} as compared to the {\bf VV} run.
At the same time, with {\bf MTS-15} we could achieve a speed-up of $\sim$7 fold.
%
It is crucial to notice that application of $\mathbf V_{\rm X}^{\rm ACE}$ operator at every SCF cycle
in place of the exact exchange operator $\mathbf V_{\rm X}$ gives a speed-up of $\sim$240 (See Table~\ref{table2}). 
However, construction of $\mathbf V_{\rm X}^{\rm ACE}$ operator, which is done only
once in every MD time step, is computationally expensive (and has the same computational 
cost of applying the exact exchange operator). 
Thus in this method, construction of $\mathbf V_{\rm X}^{\rm ACE}$ remains as the computational bottleneck. 
%



In conclusion, we presented a new scheme in using r-RESPA to perform hybrid functional based AIMD simulations with PW basis set.
This involves artificial splitting in the nuclear forces envisaged by the recently developed ACE approach.
%
%
%
%
%
Our benchmark results for liquid water show that stable and accurate MD trajectories can be obtained through this procedure. 
For the specific case of 32-water system, a computational speed-up up to 7 could be obtained.
%
%
%
%
We hope that this approach will enable us to compute long accurate AIMD trajectories at the level of hybrid DFT.
Further systematic improvement can be made to speed-up this approach, in particular the construction of ACE operator,\cite{Carnimeo_2019} and is beyond the scope of this work.
%




\begin{acknowledgments}
Authors acknowledge the HPC facility at the Indian Institute of Technology Kanpur (IITK) for the computational resources.
%
SM thanks the University Grant Commission (UGC), India, for his Ph.D. fellowship.
SM is grateful to Mr. Banshi Das (IITK) for his help in generating the power spectrum.
\end{acknowledgments}


%
\bibliography{hfx_scdm}

\begin{thebibliography}{25}%
\makeatletter
\providecommand \@ifxundefined [1]{%
 \@ifx{#1\undefined}
}%
\providecommand \@ifnum [1]{%
 \ifnum #1\expandafter \@firstoftwo
 \else \expandafter \@secondoftwo
 \fi
}%
\providecommand \@ifx [1]{%
 \ifx #1\expandafter \@firstoftwo
 \else \expandafter \@secondoftwo
 \fi
}%
\providecommand \natexlab [1]{#1}%
\providecommand \enquote  [1]{``#1''}%
\providecommand \bibnamefont  [1]{#1}%
\providecommand \bibfnamefont [1]{#1}%
\providecommand \citenamefont [1]{#1}%
\providecommand \href@noop [0]{\@secondoftwo}%
\providecommand \href [0]{\begingroup \@sanitize@url \@href}%
\providecommand \@href[1]{\@@startlink{#1}\@@href}%
\providecommand \@@href[1]{\endgroup#1\@@endlink}%
\providecommand \@sanitize@url [0]{\catcode `\\12\catcode `\$12\catcode
  `\&12\catcode `\#12\catcode `\^12\catcode `\_12\catcode `\%12\relax}%
\providecommand \@@startlink[1]{}%
\providecommand \@@endlink[0]{}%
\providecommand \url  [0]{\begingroup\@sanitize@url \@url }%
\providecommand \@url [1]{\endgroup\@href {#1}{\urlprefix }}%
\providecommand \urlprefix  [0]{URL }%
\providecommand \Eprint [0]{\href }%
\providecommand \doibase [0]{http://dx.doi.org/}%
\providecommand \selectlanguage [0]{\@gobble}%
\providecommand \bibinfo  [0]{\@secondoftwo}%
\providecommand \bibfield  [0]{\@secondoftwo}%
\providecommand \translation [1]{[#1]}%
\providecommand \BibitemOpen [0]{}%
\providecommand \bibitemStop [0]{}%
\providecommand \bibitemNoStop [0]{.\EOS\space}%
\providecommand \EOS [0]{\spacefactor3000\relax}%
\providecommand \BibitemShut  [1]{\csname bibitem#1\endcsname}%
\let\auto@bib@innerbib\@empty
\bibitem [{\citenamefont {Marx}\ and\ \citenamefont
  {Hutter}(2009)}]{marx-hutter-book}%
  \BibitemOpen
  \bibfield  {author} {\bibinfo {author} {\bibfnamefont {D.}~\bibnamefont
  {Marx}}\ and\ \bibinfo {author} {\bibfnamefont {J.}~\bibnamefont {Hutter}},\
  }\href@noop {} {\emph {\bibinfo {title} {Ab Initio Molecular Dynamics: Basic
  Theory and Advanced Methods}}}\ (\bibinfo  {publisher} {Cambridge University
  Press},\ \bibinfo {address} {Cambridge},\ \bibinfo {year} {2009})\BibitemShut
  {NoStop}%
\bibitem [{\citenamefont {Todorova}\ \emph {et~al.}(2006)\citenamefont
  {Todorova}, \citenamefont {Seitsonen}, \citenamefont {Hutter}, \citenamefont
  {Kuo},\ and\ \citenamefont {Mundy}}]{JPCB_AIMD_HFX}%
  \BibitemOpen
  \bibfield  {author} {\bibinfo {author} {\bibfnamefont {T.}~\bibnamefont
  {Todorova}}, \bibinfo {author} {\bibfnamefont {A.~P.}\ \bibnamefont
  {Seitsonen}}, \bibinfo {author} {\bibfnamefont {J.}~\bibnamefont {Hutter}},
  \bibinfo {author} {\bibfnamefont {I.-F.~W.}\ \bibnamefont {Kuo}}, \ and\
  \bibinfo {author} {\bibfnamefont {C.~J.}\ \bibnamefont {Mundy}},\ }\href
  {\doibase 10.1021/jp055127v} {\bibfield  {journal} {\bibinfo  {journal} {J.
  Phys. Chem. B}\ }\textbf {\bibinfo {volume} {110}},\ \bibinfo {pages} {3685}
  (\bibinfo {year} {2006})}\BibitemShut {NoStop}%
\bibitem [{\citenamefont {Zhang}\ \emph {et~al.}(2011)\citenamefont {Zhang},
  \citenamefont {Donadio}, \citenamefont {Gygi},\ and\ \citenamefont
  {Galli}}]{JCTC_AIMD_HFX}%
  \BibitemOpen
  \bibfield  {author} {\bibinfo {author} {\bibfnamefont {C.}~\bibnamefont
  {Zhang}}, \bibinfo {author} {\bibfnamefont {D.}~\bibnamefont {Donadio}},
  \bibinfo {author} {\bibfnamefont {F.}~\bibnamefont {Gygi}}, \ and\ \bibinfo
  {author} {\bibfnamefont {G.}~\bibnamefont {Galli}},\ }\href {\doibase
  10.1021/ct2000952} {\bibfield  {journal} {\bibinfo  {journal} {J. Chem.
  Theory Comput.}\ }\textbf {\bibinfo {volume} {7}},\ \bibinfo {pages} {1443}
  (\bibinfo {year} {2011})}\BibitemShut {NoStop}%
\bibitem [{\citenamefont {DiStasio~Jr.}\ \emph {et~al.}(2014)\citenamefont
  {DiStasio~Jr.}, \citenamefont {Santra}, \citenamefont {Li}, \citenamefont
  {Wu},\ and\ \citenamefont {Car}}]{JCP_AIMD_HFX}%
  \BibitemOpen
  \bibfield  {author} {\bibinfo {author} {\bibfnamefont {R.~A.}\ \bibnamefont
  {DiStasio~Jr.}}, \bibinfo {author} {\bibfnamefont {B.}~\bibnamefont
  {Santra}}, \bibinfo {author} {\bibfnamefont {Z.}~\bibnamefont {Li}}, \bibinfo
  {author} {\bibfnamefont {X.}~\bibnamefont {Wu}}, \ and\ \bibinfo {author}
  {\bibfnamefont {R.}~\bibnamefont {Car}},\ }\href {\doibase 10.1063/1.4893377}
  {\bibfield  {journal} {\bibinfo  {journal} {J. Chem. Phys.}\ }\textbf
  {\bibinfo {volume} {141}},\ \bibinfo {pages} {084502} (\bibinfo {year}
  {2014})}\BibitemShut {NoStop}%
\bibitem [{\citenamefont {Ambrosio}, \citenamefont {Miceli},\ and\
  \citenamefont {Pasquarello}(2016)}]{JPCB_water_hfx}%
  \BibitemOpen
  \bibfield  {author} {\bibinfo {author} {\bibfnamefont {F.}~\bibnamefont
  {Ambrosio}}, \bibinfo {author} {\bibfnamefont {G.}~\bibnamefont {Miceli}}, \
  and\ \bibinfo {author} {\bibfnamefont {A.}~\bibnamefont {Pasquarello}},\
  }\href {\doibase 10.1021/acs.jpcb.6b03876} {\bibfield  {journal} {\bibinfo
  {journal} {J. Phys. Chem. B}\ }\textbf {\bibinfo {volume} {120}},\ \bibinfo
  {pages} {7456} (\bibinfo {year} {2016})}\BibitemShut {NoStop}%
\bibitem [{\citenamefont {Tuckerman}, \citenamefont {Martyna},\ and\
  \citenamefont {Berne}(1990)}]{MTS_1}%
  \BibitemOpen
  \bibfield  {author} {\bibinfo {author} {\bibfnamefont {M.~E.}\ \bibnamefont
  {Tuckerman}}, \bibinfo {author} {\bibfnamefont {G.~J.}\ \bibnamefont
  {Martyna}}, \ and\ \bibinfo {author} {\bibfnamefont {B.~J.}\ \bibnamefont
  {Berne}},\ }\href {\doibase 10.1063/1.459140} {\bibfield  {journal} {\bibinfo
   {journal} {J. Chem. Phys.}\ }\textbf {\bibinfo {volume} {93}},\ \bibinfo
  {pages} {1287} (\bibinfo {year} {1990})}\BibitemShut {NoStop}%
\bibitem [{\citenamefont {Tuckerman}, \citenamefont {Berne},\ and\
  \citenamefont {Martyna}(1992)}]{r-RESPA}%
  \BibitemOpen
  \bibfield  {author} {\bibinfo {author} {\bibfnamefont {M.}~\bibnamefont
  {Tuckerman}}, \bibinfo {author} {\bibfnamefont {B.~J.}\ \bibnamefont
  {Berne}}, \ and\ \bibinfo {author} {\bibfnamefont {G.~J.}\ \bibnamefont
  {Martyna}},\ }\href {\doibase 10.1063/1.463137} {\bibfield  {journal}
  {\bibinfo  {journal} {J. Chem. Phys.}\ }\textbf {\bibinfo {volume} {97}},\
  \bibinfo {pages} {1990} (\bibinfo {year} {1992})}\BibitemShut {NoStop}%
\bibitem [{\citenamefont {Wu}, \citenamefont {Selloni},\ and\ \citenamefont
  {Car}(2009)}]{PRB_Car_Wannier}%
  \BibitemOpen
  \bibfield  {author} {\bibinfo {author} {\bibfnamefont {X.}~\bibnamefont
  {Wu}}, \bibinfo {author} {\bibfnamefont {A.}~\bibnamefont {Selloni}}, \ and\
  \bibinfo {author} {\bibfnamefont {R.}~\bibnamefont {Car}},\ }\href {\doibase
  10.1103/PhysRevB.79.085102} {\bibfield  {journal} {\bibinfo  {journal} {Phys.
  Rev. B}\ }\textbf {\bibinfo {volume} {79}},\ \bibinfo {pages} {085102}
  (\bibinfo {year} {2009})}\BibitemShut {NoStop}%
\bibitem [{\citenamefont {Gygi}\ and\ \citenamefont
  {Duchemin}(2013)}]{JCTC_RSB}%
  \BibitemOpen
  \bibfield  {author} {\bibinfo {author} {\bibfnamefont {F.}~\bibnamefont
  {Gygi}}\ and\ \bibinfo {author} {\bibfnamefont {I.}~\bibnamefont
  {Duchemin}},\ }\href {\doibase 10.1021/ct3007088} {\bibfield  {journal}
  {\bibinfo  {journal} {J. Chem. Theory Comput.}\ }\textbf {\bibinfo {volume}
  {9}},\ \bibinfo {pages} {582} (\bibinfo {year} {2013})}\BibitemShut {NoStop}%
\bibitem [{\citenamefont {Dawson}\ and\ \citenamefont
  {Gygi}(2015)}]{JCTC_RSB_1}%
  \BibitemOpen
  \bibfield  {author} {\bibinfo {author} {\bibfnamefont {W.}~\bibnamefont
  {Dawson}}\ and\ \bibinfo {author} {\bibfnamefont {F.}~\bibnamefont {Gygi}},\
  }\href {\doibase 10.1021/acs.jctc.5b00826} {\bibfield  {journal} {\bibinfo
  {journal} {J. Chem. Theory Comput.}\ }\textbf {\bibinfo {volume} {11}},\
  \bibinfo {pages} {4655} (\bibinfo {year} {2015})}\BibitemShut {NoStop}%
\bibitem [{\citenamefont {Ratcliff}\ \emph {et~al.}(2018)\citenamefont
  {Ratcliff}, \citenamefont {Degomme}, \citenamefont {Flores-Livas},
  \citenamefont {Goedecker},\ and\ \citenamefont {Genovese}}]{HFX_Goedecker}%
  \BibitemOpen
  \bibfield  {author} {\bibinfo {author} {\bibfnamefont {L.~E.}\ \bibnamefont
  {Ratcliff}}, \bibinfo {author} {\bibfnamefont {A.}~\bibnamefont {Degomme}},
  \bibinfo {author} {\bibfnamefont {J.~A.}\ \bibnamefont {Flores-Livas}},
  \bibinfo {author} {\bibfnamefont {S.}~\bibnamefont {Goedecker}}, \ and\
  \bibinfo {author} {\bibfnamefont {L.}~\bibnamefont {Genovese}},\ }\href
  {http://stacks.iop.org/0953-8984/30/i=9/a=095901} {\bibfield  {journal}
  {\bibinfo  {journal} {J. Phys.: Condens. Matter}\ }\textbf {\bibinfo {volume}
  {30}},\ \bibinfo {pages} {095901} (\bibinfo {year} {2018})}\BibitemShut
  {NoStop}%
\bibitem [{\citenamefont {Mandal}\ \emph {et~al.}(2018)\citenamefont {Mandal},
  \citenamefont {Debnath}, \citenamefont {Meyer},\ and\ \citenamefont
  {Nair}}]{JCP_sagar}%
  \BibitemOpen
  \bibfield  {author} {\bibinfo {author} {\bibfnamefont {S.}~\bibnamefont
  {Mandal}}, \bibinfo {author} {\bibfnamefont {J.}~\bibnamefont {Debnath}},
  \bibinfo {author} {\bibfnamefont {B.}~\bibnamefont {Meyer}}, \ and\ \bibinfo
  {author} {\bibfnamefont {N.~N.}\ \bibnamefont {Nair}},\ }\href@noop {}
  {\bibfield  {journal} {\bibinfo  {journal} {J. Chem. Phys.}\ }\textbf
  {\bibinfo {volume} {149}},\ \bibinfo {pages} {144113} (\bibinfo {year}
  {2018})}\BibitemShut {NoStop}%
\bibitem [{\citenamefont {Guidon}\ \emph {et~al.}(2008)\citenamefont {Guidon},
  \citenamefont {Schiffmann}, \citenamefont {Hutter},\ and\ \citenamefont
  {VandeVondele}}]{HFX_Hutter_JCP}%
  \BibitemOpen
  \bibfield  {author} {\bibinfo {author} {\bibfnamefont {M.}~\bibnamefont
  {Guidon}}, \bibinfo {author} {\bibfnamefont {F.}~\bibnamefont {Schiffmann}},
  \bibinfo {author} {\bibfnamefont {J.}~\bibnamefont {Hutter}}, \ and\ \bibinfo
  {author} {\bibfnamefont {J.}~\bibnamefont {VandeVondele}},\ }\href {\doibase
  10.1063/1.2931945} {\bibfield  {journal} {\bibinfo  {journal} {J. Chem.
  Phys.}\ }\textbf {\bibinfo {volume} {128}},\ \bibinfo {pages} {214104}
  (\bibinfo {year} {2008})}\BibitemShut {NoStop}%
\bibitem [{\citenamefont {Liberatore}, \citenamefont {Meli},\ and\
  \citenamefont {Rothlisberger}(2018)}]{MTS_AIMD_Ursula}%
  \BibitemOpen
  \bibfield  {author} {\bibinfo {author} {\bibfnamefont {E.}~\bibnamefont
  {Liberatore}}, \bibinfo {author} {\bibfnamefont {R.}~\bibnamefont {Meli}}, \
  and\ \bibinfo {author} {\bibfnamefont {U.}~\bibnamefont {Rothlisberger}},\
  }\href {\doibase 10.1021/acs.jctc.7b01189} {\bibfield  {journal} {\bibinfo
  {journal} {J.~Chem.~Theory Comput.}\ }\textbf {\bibinfo {volume} {14}},\
  \bibinfo {pages} {2834} (\bibinfo {year} {2018})}\BibitemShut {NoStop}%
\bibitem [{\citenamefont {Fatehi}\ and\ \citenamefont
  {Steele}(2015)}]{MTS_AIMD_Steele_3}%
  \BibitemOpen
  \bibfield  {author} {\bibinfo {author} {\bibfnamefont {S.}~\bibnamefont
  {Fatehi}}\ and\ \bibinfo {author} {\bibfnamefont {R.~P.}\ \bibnamefont
  {Steele}},\ }\href {\doibase 10.1021/ct500904x} {\bibfield  {journal}
  {\bibinfo  {journal} {J.~Chem.~Theory Comput.}\ }\textbf {\bibinfo {volume}
  {11}},\ \bibinfo {pages} {884} (\bibinfo {year} {2015})}\BibitemShut
  {NoStop}%
\bibitem [{\citenamefont {Lin}(2016)}]{ACE_Lin}%
  \BibitemOpen
  \bibfield  {author} {\bibinfo {author} {\bibfnamefont {L.}~\bibnamefont
  {Lin}},\ }\href {\doibase 10.1021/acs.jctc.6b00092} {\bibfield  {journal}
  {\bibinfo  {journal} {J.~Chem.~Theory Comput.}\ }\textbf {\bibinfo {volume}
  {12}},\ \bibinfo {pages} {2242} (\bibinfo {year} {2016})}\BibitemShut
  {NoStop}%
\bibitem [{\citenamefont {Hu}\ \emph {et~al.}(2017)\citenamefont {Hu},
  \citenamefont {Lin}, \citenamefont {Banerjee}, \citenamefont {Vecharynski},\
  and\ \citenamefont {Yang}}]{ACE_Lin_1}%
  \BibitemOpen
  \bibfield  {author} {\bibinfo {author} {\bibfnamefont {W.}~\bibnamefont
  {Hu}}, \bibinfo {author} {\bibfnamefont {L.}~\bibnamefont {Lin}}, \bibinfo
  {author} {\bibfnamefont {A.~S.}\ \bibnamefont {Banerjee}}, \bibinfo {author}
  {\bibfnamefont {E.}~\bibnamefont {Vecharynski}}, \ and\ \bibinfo {author}
  {\bibfnamefont {C.}~\bibnamefont {Yang}},\ }\href {\doibase
  10.1021/acs.jctc.6b01184} {\bibfield  {journal} {\bibinfo  {journal}
  {J.~Chem.~Theory Comput.}\ }\textbf {\bibinfo {volume} {13}},\ \bibinfo
  {pages} {1188} (\bibinfo {year} {2017})}\BibitemShut {NoStop}%
\bibitem [{\citenamefont {Chawla}\ and\ \citenamefont
  {Voth}(1998)}]{JCP_HFX_Voth}%
  \BibitemOpen
  \bibfield  {author} {\bibinfo {author} {\bibfnamefont {S.}~\bibnamefont
  {Chawla}}\ and\ \bibinfo {author} {\bibfnamefont {G.~A.}\ \bibnamefont
  {Voth}},\ }\href {\doibase 10.1063/1.476307} {\bibfield  {journal} {\bibinfo
  {journal} {J. Chem. Phys.}\ }\textbf {\bibinfo {volume} {108}},\ \bibinfo
  {pages} {4697} (\bibinfo {year} {1998})}\BibitemShut {NoStop}%
\bibitem [{\citenamefont {{Hutter et al.}}()}]{cpmd}%
  \BibitemOpen
  \bibfield  {author} {\bibinfo {author} {\bibfnamefont {J.}~\bibnamefont
  {{Hutter et al.}}},\ }\href@noop {} {\emph {\bibinfo {title} {{CPMD}: {A}n
  {A}b {I}nitio {E}lectronic {S}tructure and {M}olecular {D}ynamics
  {P}rogram}}},\ \bibinfo {note} {see {\tt http://www.cpmd.org}}\BibitemShut
  {NoStop}%
\bibitem [{\citenamefont {Adamo}\ and\ \citenamefont
  {Barone}(1999)}]{JCP_PBE0_model}%
  \BibitemOpen
  \bibfield  {author} {\bibinfo {author} {\bibfnamefont {C.}~\bibnamefont
  {Adamo}}\ and\ \bibinfo {author} {\bibfnamefont {V.}~\bibnamefont {Barone}},\
  }\href {\doibase 10.1063/1.478522} {\bibfield  {journal} {\bibinfo  {journal}
  {J. Chem. Phys.}\ }\textbf {\bibinfo {volume} {110}},\ \bibinfo {pages}
  {6158} (\bibinfo {year} {1999})}\BibitemShut {NoStop}%
\bibitem [{\citenamefont {Troullier}\ and\ \citenamefont
  {Martins}(1991)}]{PRB_TM}%
  \BibitemOpen
  \bibfield  {author} {\bibinfo {author} {\bibfnamefont {N.}~\bibnamefont
  {Troullier}}\ and\ \bibinfo {author} {\bibfnamefont {J.~L.}\ \bibnamefont
  {Martins}},\ }\href {\doibase 10.1103/PhysRevB.43.1993} {\bibfield  {journal}
  {\bibinfo  {journal} {Phys. Rev. B}\ }\textbf {\bibinfo {volume} {43}},\
  \bibinfo {pages} {1993} (\bibinfo {year} {1991})}\BibitemShut {NoStop}%
\bibitem [{\citenamefont {Martyna}, \citenamefont {Klein},\ and\ \citenamefont
  {Tuckerman}(1992)}]{NHC}%
  \BibitemOpen
  \bibfield  {author} {\bibinfo {author} {\bibfnamefont {G.~J.}\ \bibnamefont
  {Martyna}}, \bibinfo {author} {\bibfnamefont {M.~L.}\ \bibnamefont {Klein}},
  \ and\ \bibinfo {author} {\bibfnamefont {M.}~\bibnamefont {Tuckerman}},\
  }\href {\doibase 10.1063/1.463940} {\bibfield  {journal} {\bibinfo  {journal}
  {J. Chem. Phys.}\ }\textbf {\bibinfo {volume} {97}},\ \bibinfo {pages} {2635}
  (\bibinfo {year} {1992})}\BibitemShut {NoStop}%
\bibitem [{\citenamefont {Ma}, \citenamefont {Izaguirre},\ and\ \citenamefont
  {Skeel}(2003)}]{mts_resonance}%
  \BibitemOpen
  \bibfield  {author} {\bibinfo {author} {\bibfnamefont {Q.}~\bibnamefont
  {Ma}}, \bibinfo {author} {\bibfnamefont {J.}~\bibnamefont {Izaguirre}}, \
  and\ \bibinfo {author} {\bibfnamefont {R.}~\bibnamefont {Skeel}},\ }\href
  {\doibase 10.1137/S1064827501399833} {\bibfield  {journal} {\bibinfo
  {journal} {SIAM J. Sci. Comput.}\ }\textbf {\bibinfo {volume} {24}},\
  \bibinfo {pages} {1951} (\bibinfo {year} {2003})}\BibitemShut {NoStop}%
\bibitem [{\citenamefont {Kolafa}(2004)}]{JCC_ASPC}%
  \BibitemOpen
  \bibfield  {author} {\bibinfo {author} {\bibfnamefont {J.}~\bibnamefont
  {Kolafa}},\ }\href {\doibase 10.1002/jcc.10385} {\bibfield  {journal}
  {\bibinfo  {journal} {J. Comput. Chem.}\ }\textbf {\bibinfo {volume} {25}},\
  \bibinfo {pages} {335} (\bibinfo {year} {2004})}\BibitemShut {NoStop}%
\bibitem [{\citenamefont {Carnimeo}, \citenamefont {Baroni},\ and\
  \citenamefont {Giannozzi}(2019)}]{Carnimeo_2019}%
  \BibitemOpen
  \bibfield  {author} {\bibinfo {author} {\bibfnamefont {I.}~\bibnamefont
  {Carnimeo}}, \bibinfo {author} {\bibfnamefont {S.}~\bibnamefont {Baroni}}, \
  and\ \bibinfo {author} {\bibfnamefont {P.}~\bibnamefont {Giannozzi}},\ }\href
  {\doibase 10.1088/2516-1075/aaf7d4} {\bibfield  {journal} {\bibinfo
  {journal} {Electron. Struct.}\ }\textbf {\bibinfo {volume} {1}},\ \bibinfo
  {pages} {015009} (\bibinfo {year} {2019})}\BibitemShut {NoStop}%
\end{thebibliography}%

\end{document}